\documentclass[osajnl,twocolumn,showpacs,
superscriptaddress,10pt]{revtex4-1}

\usepackage{graphicx}
\usepackage{bm}
\usepackage{textcomp}

\begin{document}

\title{Photonic bands and defect modes in metallo-dielectric photonic crystal slabs} 

\author{Simone Zanotto}
\email{simone.zanotto@sns.it}
\affiliation{NEST, Istituto Nanoscienze - CNR and Scuola Normale Superiore, P.za S. Silvestro 12, I-56127 Pisa, Italy}

\author{Giorgio Biasiol} 
\affiliation{CNR-IOM, Laboratorio TASC, Area Science Park, I-34149 Trieste, Italy}

\author{Lucia Sorba}
\author{Alessandro Tredicucci}
\affiliation{NEST, Istituto Nanoscienze - CNR and Scuola Normale Superiore, P.za S. Silvestro 12, I-56127 Pisa, Italy}

\date{\today}   

\begin{abstract}
Photonic components based on structured metallic elements show great potential for device applications where field enhancement and confinement of the radiation on a subwavelength scale is required. In this paper we report a detailed study of a prototypical metallo-dielectric photonic structure, where features well known in the world of dielectric photonic crystals, like band gaps and defect modes, are exported to the metallic counterpart, with interesting applications to infrared science and technology, as for instance in quantum well infrared photodetectors, narrow-band spectral filters, and tailorable thermal emitters.
\end{abstract}


\maketitle 

The idea of reproducing the naturally occurring periodic arrangement of atoms in solid crystals by settling ordered arrays of scattering elements on the light wavelength scale gave origin to the extremely fruitful concept of photonic crystal\cite{YablonovitchPRL1987}. Despite the full maturity of this research field, new possibilities keep emerging, regarding the ability of spatially confining otherwise freely propagating photons\cite{HsuNat2013, MarinicaPRL2008}. In original proposals and traditional implementations, photonic crystals relied on dielectric media, mainly owing to the natural interconnection with semiconductor science and technology. However, moving to metallic photonic structures, new physics was developed and technological milestones were set. Starting from the discovery of extraordinary transmission\cite{EbbesenNat1998}, researchers explored the world of periodic arrangements of metallic elements, where the concepts of metamaterials and plasmonics naturally appear. 

Hybrid metallo-dielectric photonic devices attract attention because the interface between those two classes of materials naturally supports an intense local electric field. This is of clear relevance when the goal is to enhance non-linear effects\cite{KleinPRB2005}, but also in the realm of linear physics, where the existence of a maximum of the field close to an interface enables the technology for a whole class of photonic devices, like terahertz quantum cascade lasers (QCLs). It is indeed in connection with QCLs and other intersubband devices that certain concepts, like distributed feedback lasers and photonic crystal resonators, have been exported to metallo-dielectric structures\cite{ZanottoAPL2010, MahlerAPL2004, ChassagneuxNat2009}.

In this paper we theoretically and experimentally analyze a simple one-dimensional metallo-dielectric photonic slab structure, where two key features of photonic crystals --- band gap and defect modes --- clearly appear. The defect mode is a high Q-factor resonance, whose linewidth and angular acceptance can be tuned at will within wide ranges. The observed phenomenology is general and can be extrapolated to devices working in different wavelength ranges\cite{ChristPRL2003, ZentgrafPRB2006}; furthermore, small modifications to the studied prototypical passive component could directly lead to operating devices for the infrared spectral range. 
Thanks to the anticipated resonance tunability, metallo-dielectric photonic crystal defects can be very useful for the development of guided-mode resonance filters (GMRF)\cite{SakatOL2011, SakatOE2012} and thermal emitters\cite{LiuPRL2011} exceeding state-of-the-art peformance. In addition, the presence of a metal, which helps electrical contacting, and the absence of active region removal upon etching, could be of technological relevance in the design of efficient quantum-well infrared photodetectors (QWIPs)\cite{KalchmairAPL2011, SchartnerOE2008}.

\begin{figure}[b]
	\centering
		\includegraphics{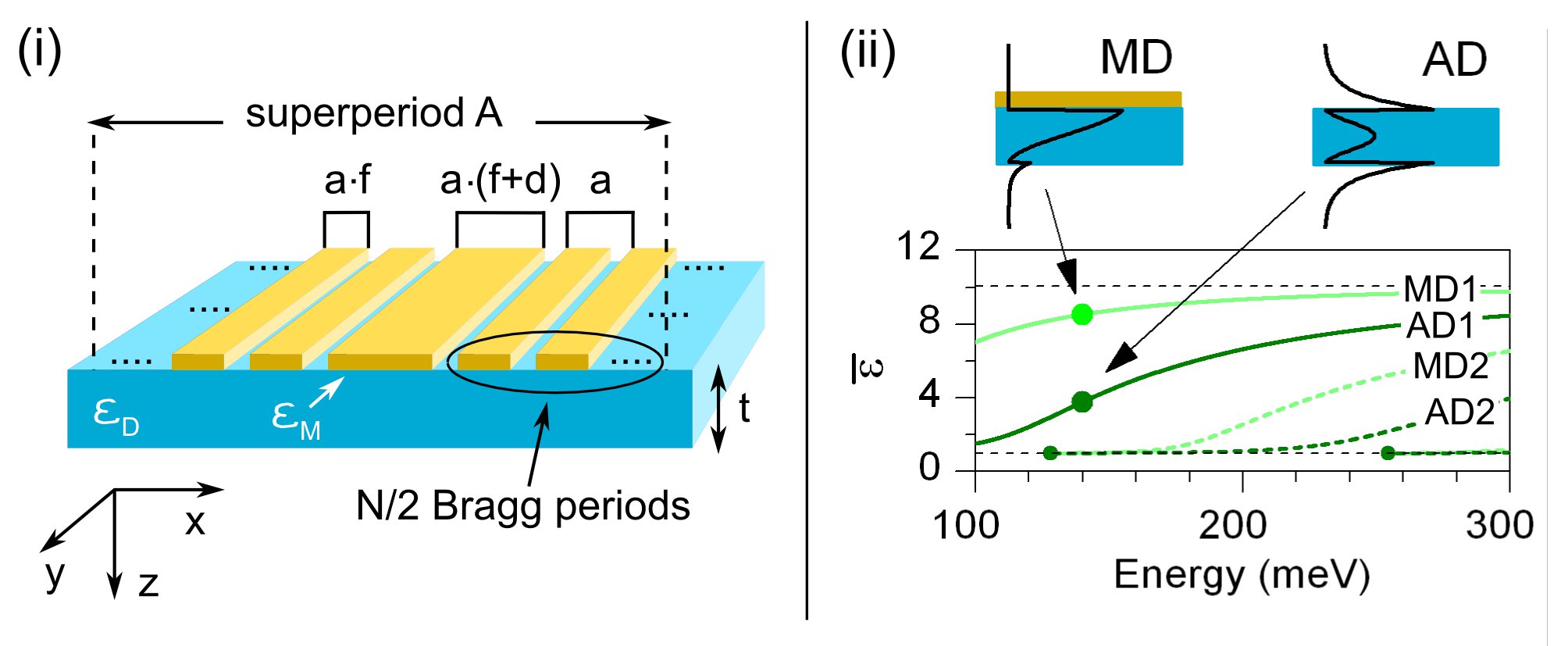}
	\caption{(Color online) (i) Schematic representation of one supercell of the photonic crystal slab: a dielectric membrane with permittivity $\varepsilon_D > 0$ patterned with metal stripes with permittivity $\varepsilon_M < 0$. (ii) Guided modes supported by the metallo-dielectric (MD) and air-dielectric (AD) planar waveguides involved in the analysis of the photonic crystal response. The insets represent the $E_z$ field profile at the selected photon energy $E = 140$ meV, where only the first order modes are relevant. The horizontal dashed lines correspond to air and slab dielectric constants.}
	\label{Figure1}
\end{figure}

The device under study is schematically represented in Fig.\ \ref{Figure1} (i). 
It consists of a dielectric membrane patterned with thin metallic stripes arranged in a supercell scheme, where a central stripe (defect) is surrounded by a region containing a set of $N$ equispaced stripes acting as a photonic Bragg mirror. We indicate with $a$ and $f$ the lattice constant and ``duty cycle'' of the Bragg mirror, while $d$ is connected to the defect width; all the stripes are spaced by an air gap of width $a \cdot (1-f)$. If $d = 0$ there is no defect, and the whole structure is periodic with the Bragg mirror period $a$; if $d \neq 0$ the structure has a period $A = a \cdot (N+1+d)$. Having in mind possible future applications to active intersubband devices, the dielectric membrane consists of a $\mathrm{GaAs}/\mathrm{Al}_{0.33}\mathrm{Ga}_{0.67}\mathrm{As}$ superlattice similar to that reported in Ref.\ \onlinecite{ZanottoPRB2012R}, where the fabrication details can also be retrieved. Since the heterostructure is here undoped, its role boils down to an effective dielectric of permittivity $\varepsilon_{D} = 10.05$. The thickness is $t = 1600\ \mathrm{nm}$.

In the photonic crystal slab philosophy, the working principle of the device can be interpreted by first considering the confinement along the $z$ direction\cite{GeracePRE2004}. In this sense, the membrane is an alternance of (air)/metal/dielectric/air (MD) and air/dielectric/air (AD) slab waveguides, each of them characterized by a set of guided modes with effective dielectric constants $\bar{\varepsilon}$ [Fig.\ \ref{Figure1} (ii)]\footnote{The presence of air above the metal can be neglected since the penetration depth in the metal is smaller than its thickness. In this paper we assume $\varepsilon_{M} = -4000$ with no imaginary part}. Referring again to the link with intersubband devices, we focus here on the sole TM polarization. At the target photon energy of 140 meV, both MD and AD waveguides are substantially single-moded, and the contrast between their effective permittivities $\bar{\varepsilon}_{\mathrm{MD1}}$ and $\bar{\varepsilon}_{\mathrm{AD1}}$ is quite large. Hence, we expect that the Bragg mirror regions will actually exhibit a stop band for a wave travelling along the $x$ direction, and in the forthcoming we will show that this is the mechanism which enables the promotion of a confined defect mode.

Before focusing on a thorough discussion about the opening of a photonic gap and the consequent study of defect modes, we present in Fig.\ 2
\begin{figure}[tb]
	\centering
		\includegraphics{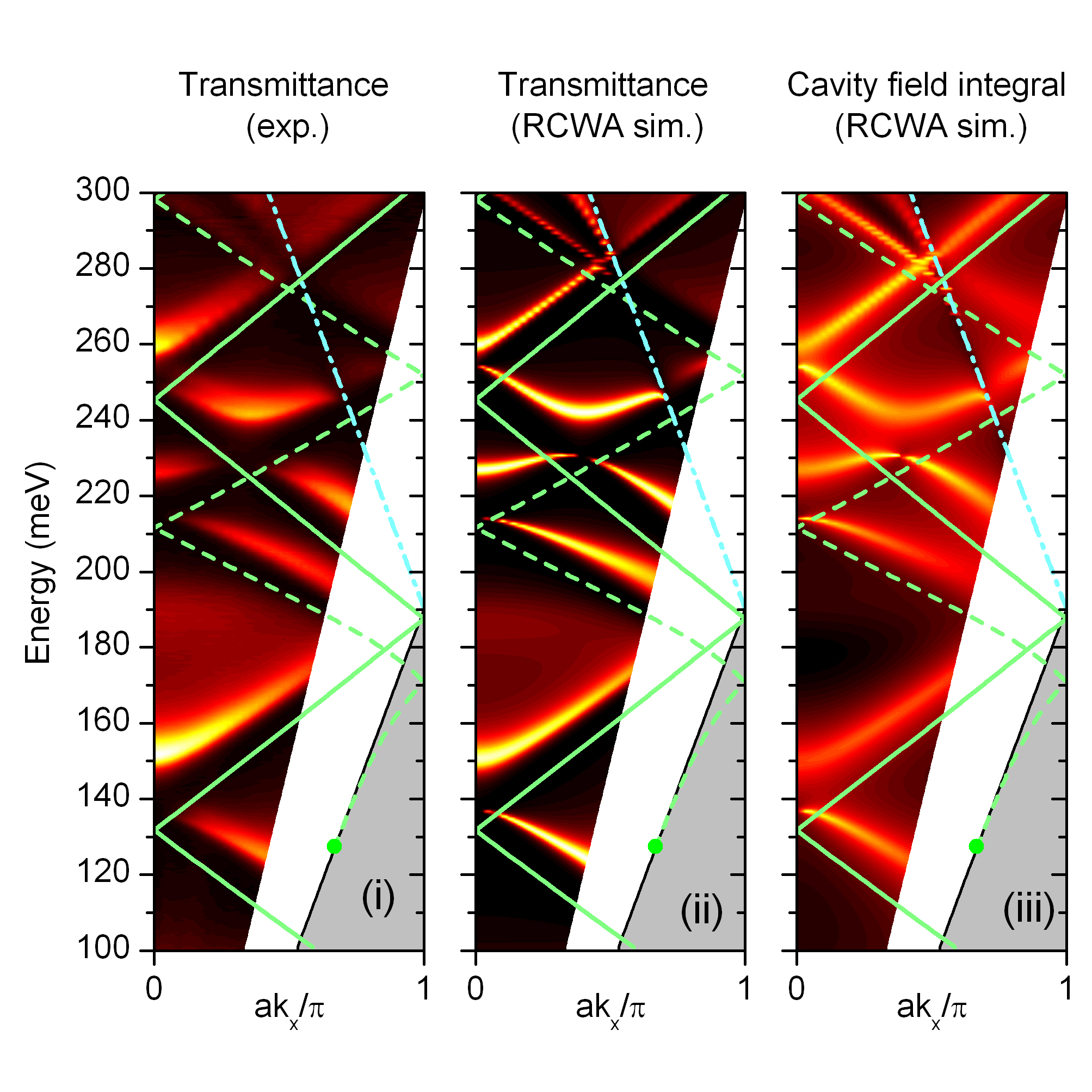}
	\caption{(Color online) Photonic bands of the photonic crystal slab without defect, emerging from transmittance and resonant intracavity field enhancement. The color maps are coded as following: black, low; yellow, high. Green continuous (dashed) lines: folding of the first (second) mode of the MD planar waveguide in the first Brillouin zone; cyan dash-dot line: $1^{\circ}$ order diffraction cone in air. White areas correspond to incidence angles greater than $40^{\circ}$ (experimental limit); gray area corresponds to points below the air light cone.}
\end{figure} 
a set of data enforcing the relevance of the MD guided modes in the optical properties of device fabricated without defect ($d = 0$), and with a large metal duty cycle ($f = 80\%$). The motivation of the latter choice lies in the intended application for QWIPs and GMRFs: a large metal fraction leads to a large overlap of the field with the dielectric, and induces clear spectral peaks in the membrane transmittance. Moreover, values close to $80\%$ maximise the photonic bandgap, enabling the implementation of defect modes\footnote{The analysis of the $f$-dependence of spectral features and bandgap is not immediate, and the interested reader can refer to Ref.\ \onlinecite{FanOE2006, ZanottoPRB2012} for further details.}. 
We probed the eigenmodes of the PhC slab lying above the light line (\textit{guided mode resonances}) by means of angle-resolved Fourier-transform spectrometry. The incidence plane was $xz$ and the light beam, set to an incidence angle $\theta$ with respect to the normal to the sample surface, was $p-$polarized. By plotting the angularly-resolved transmittance spectra in the $E$-$k_x$ plane [Fig.\ 2 (i)], it can be noticed that the spectral features closely follow the trend of the MD guided modes folded in the first Brillouin zone, i.e., the points of the $E$-$k_x$ plane which fulfil
\begin{equation}
 k_0 \sqrt{\bar{\varepsilon}_{\mathrm{MD}}} = |k_x + m g|
\label{band_folding}
\end{equation}
where $k_0 \sqrt{\bar{\varepsilon}_{\mathrm{MD}}}$ is the MD waveguide in-plane wavevector, $k_x = k_0 \sin{\theta}$ is the in-plane component of the incident light wavevector, $m$ is an integer and $g = 2 \pi/a$ is the reciprocal lattice unit vector; $k_0  = E/\hbar c$ is the light vavevector in vacuum. In essence, the incoming light couples to the MD guided mode(s) via the lattice periodicity, provided that the resonance condition (\ref{band_folding}) holds. The second MD mode is required to account for the features above 180 meV. 

In spite of its simplicity, this band folding scheme -- which is substantially a zero-order perturbation theory -- does not give insights into how the guided mode resonance couples to the far-field, i.e., into the resonance linewidth and the spectral contrast. These features can instead be quantified by relying on the rigorous coupled-wave analysis (RCWA)\cite{GranetJOSAA1996}, a numerical method which solves the full Maxwell equations given the geometry of a periodic multilayered medium. Employing the sample nominal parameters, we obtained the transmittance plot of Fig.\ 2 (ii), in good agreement with the experimental data. The simulated spectra only show slightly narrower lineshapes and a larger contrast of certain photonic bands; this could be attributed either to the non-planarity of the fabricated membrane, which shows fluctuations of $\simeq 5^{\circ}$ due to some strain in the MBE-grown $\mathrm{GaAs}/\mathrm{Al}_{0.33}\mathrm{Ga}_{0.67}\mathrm{As}$ superlattice, and to imperfections in the lithographic process. Having checked that the experimentally observed transmittance peaks correspond to a theoretical prediction, we now argue that they actually occur when a resonant field inside the membrane is excited. This can be easily stated relying on the RCWA: since it gives direct access to the local fields inside the PhC slab, the color map of Fig.\ 2 (iii) is readily obtained. In substance, a transmission peak occurs when the photon dwells for a time inside the membrane; the narrower the peak, the larger the dwelling time and the local field. In other words, the transmittance peak is the signature of a leaky eigenmode of the photonic crystal slab\cite{FanJOSAA2003}. 

In the pictures given above the transmittance peaks are related to the excitation of MD guided modes and to the photonic crystal slab eigenmodes, but no direct insight in the mechanism leading to the bandgap opening is gained. Such an insight can be obtained by introducing a simple one-dimensional photonic crystal model, where the photonic slab is replaced by a stack of homogeneous effective dielectric layer, as represented in the top panels of Fig.\ 3. This model goes beyond the zero order perturbation theory (Eq.\ \ref{band_folding}), but the computational effort is much less than that required for RCWA -- which, in case of large supercell calculations, and parametric sweeps, can become impractical.
\begin{figure}[ht]
	\centering
		\includegraphics{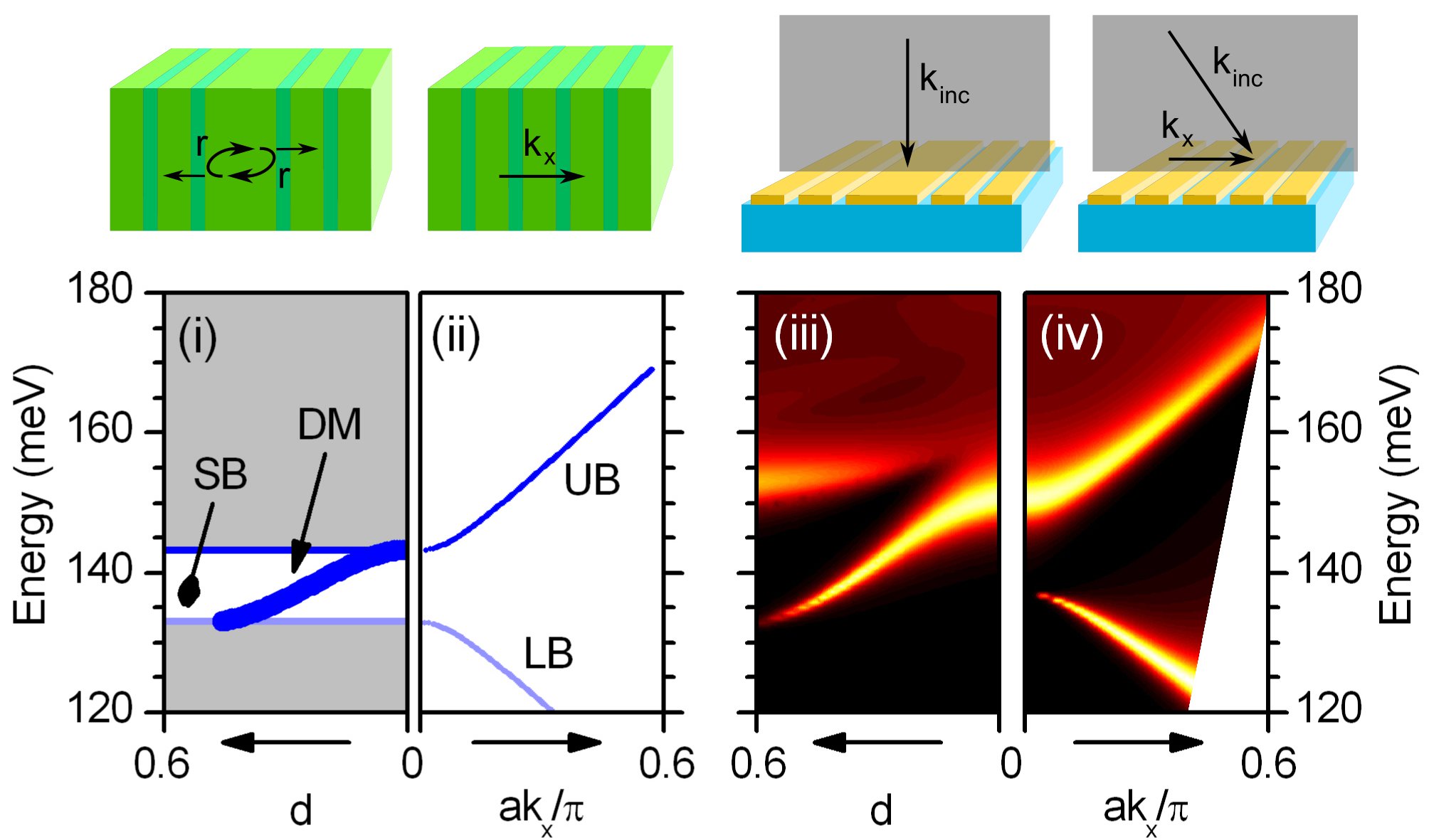}
	\caption{(Color online) Photonic bandgap and defect mode promotion. A simplified 1-d ideal photonic crystal model is proposed (i, ii) and confirmed by rigorous coupled-wave analysis transmittance calculations (iii, iv). UB (LB): upper (lower) branches; SB: stop-band; DM: defect mode.}
	\label{Figure3}
\end{figure} 
In essence, the PhC slab without defect is interpreted as a Bragg mirror constituted by layers of thicknesses $a \cdot f$ and $a \cdot (1-f)$, and permittivities $\bar{\varepsilon}_{\mathrm{MD1}}$ and $\bar{\varepsilon}_{\mathrm{AD1}}$. The band structure, calculated for the gap near $140\ \mathrm{mev}$ with the transfer-matrix method\cite{Yariv}, is reported in Fig.\ 3 (ii), and shows the second-order stop band separating the upper and lower branches. This band structure should be compared to the actual PhC slab transmission reported in Fig.\ 3 (iv): despite the non complete agreement, the 1-d PhC approach clearly predicts a gap in the correct energy range. The quantitative mismatch should be attributed to the presence of MD/AD modes other than the fundamental one, to the mode profile difference between MD and AD modes, and to the presence of radiative channels. It is indeed the coupling with the radiative modes that is a point of special relevance, since this is the mechanism that will enable the operation of actual devices. This aspect, that could be treated analytically within a perturbation theory\cite{GeracePRE2004}, is here analyzed numerically by observing the behaviour of the transmittance predicted by RCWA. In our case, at the zone center the UB (LB) is radiative (non-radiative). This is not a general phenomenon, since the role of the UB and LB can be exchanged\cite{FanOE2006}; what can be asserted is that, because of the symmetry $x \leftrightarrow -x$ in the unit cell, one of the two band edges is radiative and the other is not. 

Moving now to the search for a defect mode within the gap, in the 1-d model the condition is substantially to require constructive interference after a round-trip\cite{Lourtioz}:
\begin{equation}
 2 \mathrm{arg}(r) + 2 k_{\mathrm{def}} L_{\mathrm{def}} = 2 m \pi
\label{defect_equation}
\end{equation}
where $\mathrm{arg}(r)$ is the phase of the reflection amplitude of the (semi-infinite) DBR; $k_{\mathrm{def}} = k_0 \sqrt{\bar{\varepsilon}_{\mathrm{MD1}}}$ is the wavevector in the defect effective medium, $L_{\mathrm{def}} = a \cdot (f+d) $ is the defect thickness and $m$ is an integer. The solution of this equation as a function of $d$ is reported in Fig.\ 3 (i): the defect mode is promoted from the upper branch as $d$ increases, and eventually disappears at $d \simeq 0.5$ collapsing in the lower branch. Whether the defect mode has a radiative coupling or not is answered by RCWA [Fig.\ 3 (iii)]: at normal incidence, its spectral feature is a transmission peak continuously evolving from the bottom of the upper branch. The radiative nature of the defect mode, and the possibility to continuously tune its linewidth by keeping a large contrast in transmission, makes it appealing in view of applications. For QWIPs, the matching of radiative- and absorption-induced decay rates is a key issue (\textit{critical coupling}), and this tuning mechanism is complementary to that reported in Ref.\ \onlinecite{ManceauAPL2013}. For GMRFs, a fully-contrasted, isolated transmission feature of tunable linewidth is the essence of the device. Furthermore, the high Q-factor attainable\footnote{With these structures, a Q-factor as large as 100 can be attained without losing coupling efficiency. While not a large value in absolute terms, it is among the largest observed in mid-infrared photonic devices.}, directly connected to intracavity field enhancement\cite{Yariv}, is particularly appealing in the development of efficient two-photon detectors\cite{SchneiderAPL2008}, or, more in general, of nonlinear devices. 

While Equation (\ref{defect_equation}) describes a defect embedded between \textit{infinite} Bragg mirrors, practical implementations -- either in RCWA and in experiments -- rely on a finite number of Bragg periods N. 
In particular, the spectra of Fig.\ 3 (iii) correspond to structures with $N = 6$, a value above which the defect mode does not significantly change neither frequency nor linewidth. On the contrary, below that value the number of Bragg periods has an influence on the photonic modes, as we observed by analyzing the band structures of a set of fabricated samples which differ by $N$ at fixed $d$. The value $d = 0.4$ is chosen, since it allows for a narrow transmission linewidth while not being too close to the limit value $d \simeq 0.5$. In the top panels of Fig.\ 4 
\begin{figure*}[t]
	\centering
		\includegraphics{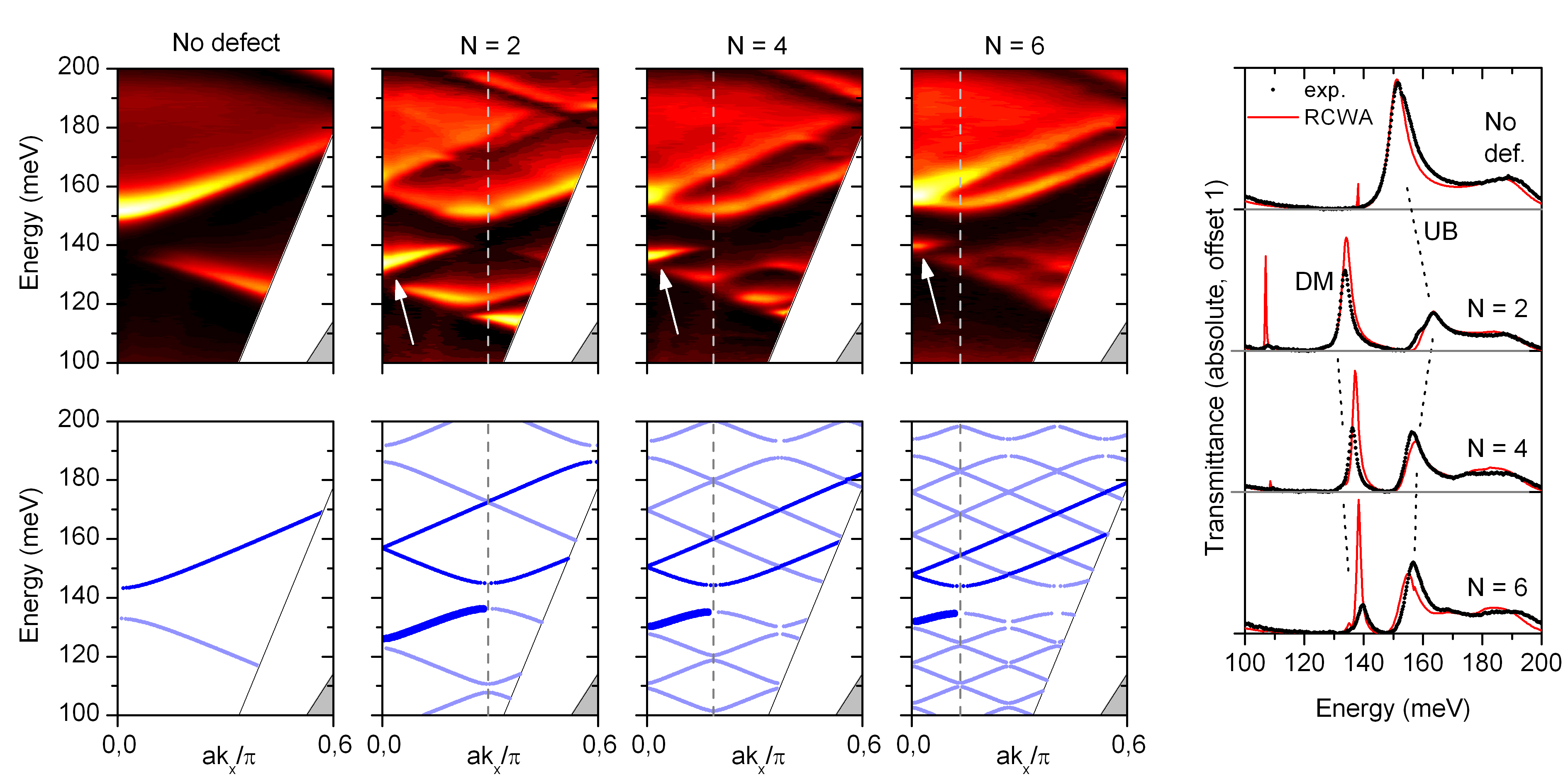}
	\caption{(Color online) (Upper panels) Fingerprints of the defect mode in the photonic bandstructure from transmittance measurements on four samples, without defect or with defect and increasing number $N$ of Bragg mirror periods. The white arrow shows the defect mode. (Lower panels) Photonic bands calculated in the 1-d ideal photonic crystal model, with the optically active defect mode evidenced as a thicker line spanning the first Brillouin zone of the superlattice. The defect mode extends until the boundary of the Brillouin zone of the supercell, evidenced as a dashed grey line. (Right panel) Transmittance spectra at normal incidence ($k_x = 0$), with the defect mode (DM) and the upper branch (UB) in evidence.}
	\label{Figure4}
\end{figure*} 
we report the measured band structures (transmittance spectra) of four samples, starting from the sample with no defect (already analyzed above) and going through $N = 2, 4, 6$; an interpretation in terms of the 1-d ideal photonic crystal model is given in the lower panels. The upper band of the photonic crystal without defect develops substantially in two radiative bands (evidenced in dark blue in the model), while the lower band develops in a manifold of features. The defect mode, highlighted by an arrow in the transmittance maps, corresponds to the band evidenced with a thicker line in the model. Its radiative nature at the $k_x = 0$ point (normal incidence) is lost upon reaching the boundary of the Brillouin zone corresponding to the supercell period ($k_x = \pi / A$, indicated with a vertical dashed line). Hence, the choice of the number of Bragg periods can be employed as a tuning tool for targetting a given angular acceptance of the device: this may have an impact on the engineering of direction-selective thermal emitters\footnote{The proposed analysis disregards the coupling with light propagating out of the $xz$ plane. It can be argued that, by implementing a defect mode in a square lattice, full control of the directionality around normal incidence can be attained.}. Finally, in the right panel of Fig.\ 4 a quantitative comparison between experiment (at normal incidence) and RCWA is proposed. Despite the non ideal experimental visibility of the defect mode\footnote{As for the photonic resonances of the sample without defect, this is attributed to sample non-planarity and lithograpic imperfections.}, a good overall agreement enforces the interpretation of the observed features in terms of localized defect modes. 

In conclusion, we performed a full study of a metallo-dielectric photonic crystal slab, where general features of photonic structures like stop bands and defect modes have been observed. The results have been interpreted at different theoretical levels, giving insights in the physics behind the eigenmode structure as well as in the design potentials for actual devices like linear and non-linear quantum well infrared photodetectors, guided mode resonance filters, and tailorable thermal emitters. 

The authors gratefully acknowledge Riccardo Degl'Innocenti who participated in stimulating discussions. This work was supported by the Italian Ministry of Economic Development through the ICE-CRUI project ``Teragraph'', and by the European Research Council through the advanced grant ``SoulMan''.


\begin{thebibliography}{10}
\newcommand{\enquote}[1]{``#1''}

\bibitem{YablonovitchPRL1987}
E.~Yablonovitch, \enquote{Inhibited spontaneous emission in solid-state physics
  and electronics,} Phys. Rev. Lett. \textbf{58}, 2059--2062 (1987).

\bibitem{HsuNat2013}
C.~W. Hsu, B.~Zhen, J.~Lee, S.-L. Chua, S.~G. Johnson, J.~D. Joannopoulos, and
  M.~Soljačić, \enquote{Observation of trapped light within the radiation
  continuum,} Nature \textbf{499}, 188--191 (2013).

\bibitem{MarinicaPRL2008}
D.~C. Marinica, A.~G. Borisov, and S.~V. Shabanov, \enquote{Bound states in the
  continuum in photonics,} Phys. Rev. Lett. \textbf{100}, 183902 (2008).

\bibitem{EbbesenNat1998}
T.~W. Ebbesen, H.~J. Lezec, H.~F. Ghaemi, T.~Thio, and P.~A. Wolff,
  \enquote{Extraordinary optical transmission through sub-wavelength hole
  arrays,} Nature \textbf{391}, 667--669 (1998).

\bibitem{KleinPRB2005}
M.~W. Klein, T.~Tritschler, M.~Wegener, and S.~Linden, \enquote{Lineshape of
  harmonic generation by metallic nanoparticles and metallic photonic crystal
  slabs,} Phys. Rev. B \textbf{72}, 115113 (2005).

\bibitem{ZanottoAPL2010}
S.~Zanotto, G.~Biasiol, R.~Degl’Innocenti, L.~Sorba, and A.~Tredicucci,
  \enquote{{Intersubband polaritons in a one-dimensional surface plasmon
  photonic crystal},} Applied Physics Letters \textbf{97}, 231123 (2010).

\bibitem{MahlerAPL2004}
L.~Mahler, R.~Kohler, A.~Tredicucci, F.~Beltram, H.~E. Beere, E.~H. Linfield,
  D.~A. Ritchie, and A.~G. Davies, \enquote{Single-mode operation of terahertz
  quantum cascade lasers with distributed feedback resonators,} Applied Physics
  Letters \textbf{84}, 5446--5448 (2004).

\bibitem{ChassagneuxNat2009}
Y.~Chassagneux, R.~Colombelli, W.~Maineult, S.~Barbieri, H.~E. Beere, D.~A.
  Ritchie, S.~P. Khanna, E.~H. Linfield, and A.~G. Davies,
  \enquote{Electrically pumped photonic-crystal terahertz lasers controlled by
  boundary conditions,} Nature \textbf{457}, 174--178 (2009).

\bibitem{ChristPRL2003}
A.~Christ, S.~G. Tikhodeev, N.~A. Gippius, J.~Kuhl, and H.~Giessen,
  \enquote{Waveguide-plasmon polaritons: Strong coupling of photonic and
  electronic resonances in a metallic photonic crystal slab,} Phys. Rev. Lett.
  \textbf{91}, 183901 (2003).

\bibitem{ZentgrafPRB2006}
T.~Zentgraf, A.~Christ, J.~Kuhl, N.~A. Gippius, S.~G. Tikhodeev, D.~Nau, and
  H.~Giessen, \enquote{Metallodielectric photonic crystal superlattices:
  Influence of periodic defects on transmission properties,} Phys. Rev. B
  \textbf{73}, 115103 (2006).

\bibitem{SakatOL2011}
E.~Sakat, G.~Vincent, P.~Ghenuche, N.~Bardou, S.~Collin, F.~Pardo, J.-L.
  Pelouard, and R.~Ha\"{i}dar, \enquote{Guided mode resonance in subwavelength
  metallodielectric free-standing grating for bandpass filtering,} Opt. Lett.
  \textbf{36}, 3054--3056 (2011).

\bibitem{SakatOE2012}
E.~Sakat, G.~Vincent, P.~Ghenuche, N.~Bardou, C.~Dupuis, S.~Collin, F.~Pardo,
  R.~Ha\"{i}dar, and J.-L. Pelouard, \enquote{Free-standing guided-mode
  resonance band-pass filters: from 1d to 2d structures,} Opt. Express
  \textbf{20}, 13082--13090 (2012).

\bibitem{LiuPRL2011}
X.~Liu, T.~Tyler, T.~Starr, A.~F. Starr, N.~M. Jokerst, and W.~J. Padilla,
  \enquote{Taming the blackbody with infrared metamaterials as selective
  thermal emitters,} Phys. Rev. Lett. \textbf{107}, 045901 (2011).

\bibitem{KalchmairAPL2011}
S.~Kalchmair, H.~Detz, G.~D. Cole, A.~M. Andrews, P.~Klang, M.~Nobile,
  R.~Gansch, C.~Ostermaier, W.~Schrenk, and G.~Strasser, \enquote{Photonic
  crystal slab quantum well infrared photodetector,} Applied Physics Letters
  \textbf{98}, 011105 (2011).

\bibitem{SchartnerOE2008}
S.~Schartner, M.~Nobile, W.~Schrenk, A.~M. Andrews, P.~Klang, and G.~Strasser,
  \enquote{Photocurrent response from photonic crystal defect modes,} Opt.
  Express \textbf{16}, 4797--4803 (2008).

\bibitem{ZanottoPRB2012R}
S.~Zanotto, R.~Degl'Innocenti, J.-H. Xu, L.~Sorba, A.~Tredicucci, and
  G.~Biasiol, \enquote{Ultrafast optical bleaching of intersubband cavity
  polaritons,} Phys. Rev. B \textbf{86}, 201302 (2012).

\bibitem{GeracePRE2004}
D.~Gerace and L.~C. Andreani, \enquote{Gap maps and intrinsic diffraction
  losses in one-dimensional photonic crystal slabs,} Phys. Rev. E \textbf{69},
  056603 (2004).

\bibitem{Note1}
The presence of air above the metal can be neglected since the penetration
  depth in the metal is smaller than its thickness. In this paper we assume
  $\varepsilon _{M} = -4000$ with no imaginary part.

\bibitem{Note2}
The analysis of the $f$-dependence of spectral features and bandgap is not
  immediate, and the interested reader can refer to Ref.\ 23 and 30 for further details.

\bibitem{GranetJOSAA1996}
G.~Granet and B.~Guizal, \enquote{Efficient implementation of the coupled-wave
  method for metallic lamellar gratings in tm polarization,} J. Opt. Soc. Am. A
  \textbf{13}, 1019--1023 (1996).

\bibitem{FanJOSAA2003}
S.~Fan, W.~Suh, and J.~D. Joannopoulos, \enquote{Temporal coupled-mode theory
  for the Fano resonance in optical resonators,} J. Opt. Soc. Am. A
  \textbf{20}, 569--572 (2003).

\bibitem{Yariv}
A.~Yariv and P.~Yeh, \emph{Photonics: Optical Electronics in Modern
  Communication} (Oxford University Press, 2007).

\bibitem{FanOE2006}
J.~A. Fan, M.~A. Belkin, F.~Capasso, S.~Khanna, M.~Lachab, A.~G. Davies, and E.~H.
  Linfield, \enquote{Surface emitting terahertz quantum cascade
  laser with a double-metal waveguide,} Opt. Express \textbf{14}, 11672--11680
  (2006).

\bibitem{Lourtioz}
J.~M. Lourtioz, H.~Benisty, V.~Berger, J.-M. Gerard, and D.~Maystre,
  \emph{Photonic Crystals: Towards Nanoscale Photonic Devices} (Springer,
  2008).

\bibitem{ManceauAPL2013}
J.-M. Manceau, S.~Zanotto, and R.~Colombelli, \enquote{Optical critical coupling into highly confining metal-insulator-metal resonators,} Applied Physics Letters \textbf{103}, 091110 (2013).

\bibitem{Note3}
With these structures, a Q-factor as large as 100 can be attained without
  losing coupling efficiency. While not a large value in absolute terms, it is
  among the largest observed in mid-infrared photonic devices.

\bibitem{SchneiderAPL2008}
H.~Schneider, H.~C. Liu, S.~Winnerl, O.~Drachenko, M.~Helm, and J.~Faist,
  \enquote{Room-temperature midinfrared two-photon photodetector,} Applied
  Physics Letters \textbf{93}, 101114 (2008).

\bibitem{Note4}
The proposed analysis disregards the coupling with light propagating out of the
  $xz$ plane. It can be argued that, by implementing a defect mode in a square
  lattice, full control of the directionality around normal incidence can be
  attained.

\bibitem{Note5}
As for the photonic resonances of the sample without defect, this is attributed
  to sample non-planarity and lithograpic imperfections.

\bibitem{ZanottoPRB2012}
S.~Zanotto, R.~Degl'Innocenti, L.~Sorba, A.~Tredicucci, and G.~Biasiol,
  \enquote{Analysis of line shapes and strong coupling with intersubband
  transitions in one-dimensional metallodielectric photonic crystal slabs,}
  Phys. Rev. B \textbf{85}, 035307 (2012).

\end{thebibliography}
\end{document}